\documentclass[letter, 10 pt, conference]{IEEEconf}  % Comment this line out
\IEEEoverridecommandlockouts                              % This command is only
\usepackage{lineno,hyperref}
\usepackage{xcolor}
\usepackage{subcaption}
\usepackage{graphicx}
\usepackage{mathtools}
\usepackage{afterpage}
\usepackage{tikz}
\usepackage{amsmath}
\usepackage{amsfonts}
\usepackage{siunitx}
\usetikzlibrary{patterns,decorations.pathreplacing}
\modulolinenumbers[5]

\newtheorem{remark}{\textbf{Remark}}

\newcommand{\mr}[1]{\mathrm{#1}}
\newcommand{\tran}{^{\mkern-1.5mu\mathsf{T}}}

\begin{document}

% \begin{frontmatter}

\title{\LARGE \bf
  Practical Guidelines for Data-driven Identification \\of Lifted Linear Predictors for Control
  % \blue{(of dynamical systems)}
}

% \corref{corr_author}

\author{Loi Do, Adam Uchytil, and Zdeněk Hurák%
\thanks{This work was co-funded by the European Union under the project ROBOPROX (reg. no. CZ.02.01.01/00/22\_008/0004590).}% <-this % stops a space
\thanks{Loi Do, Adam Uchytil, and Zdeněk Hurák are with Faculty of Electrical Engineering, Czech Technical University in Prague
        {\tt\small \{doloi, uchytada, hurak\}[at]fel.cvut.cz}}%        
}

%\author[mymainaddress]{}
%\author[mymainaddress,mysecondaryaddress]{Czech Technical University in Prague}

%s\author[mysecondaryaddress]{Global Customer Service\corref{correspondingauthor}}

%\address[mymainaddress]{Karlovo Náměstí 13, 120 00 Prague 2, Czech Republic}
% \address{Karlovo Náměstí 13, 120 00 Prague 2, Czech Republic}
% \newcommand{\zuzo}{ŽůžO}

\maketitle
\thispagestyle{empty}
\pagestyle{empty}

\begin{abstract}

  Lifted linear predictor (LLP) is an artificial linear dynamical system designed to predict trajectories of a generally nonlinear dynamical system based on the current state (or measurements) and the input.
  The main benefit of the LLP is its potential ability to capture the nonlinear system's dynamics with precision superior to other linearization techniques, such as local linearization about the operation point.
  The idea of lifting is supported by the theory of Koopman Operators. 
  For LLP identification, we focus on the data-driven method based on the extended dynamic mode decomposition (EDMD) algorithm.
  However, while the EDMD algorithm presents an extremely simple and efficient way to obtain the LLP, it can also yield poor results.
  In this paper, we present some less intuitive practical guidelines for data-driven identification of the LLPs, aiming at improving usability of LLPs for designing control.
  We support the guidelines with two motivating examples.
  % as well as with references to relevant works.
  The implementation of the examples are shared on a public repository.
\end{abstract}


% ==============================================================================================
%==============================================================================================
%==============================================================================================

\section{Introduction}

%========== V1 ==========
% Designing a controller for a nonlinear system typically requires a model of its dynamics.
% Additionally, it is beneficial to obtain a \textit{linear} model which enables using linear control systems designs.
% =======================

One approach to control design for nonlinear systems is to use linear control system methods.
To this end, a linear model of the system's dynamics is required.
There are many ways to obtain a linear model, but the primary issue with such linear models is their small validity region.

% Linear methods for control synthesis necessitate linear representation of the system's, generally nonlinear, dynamics.
% Having an accurate linear representation of a nonlinear system enables using linear methods for control synthesis.
% Designing a control for a system 
% Acquiring an accurate model of a dynamical system with is an essential step in designing a control 
% Having an accurate model of a dynamical system with is a central  for control engineering.
% with unknown dynamics
% Many classical data-driven methods.
% % A classical option, when the system's dynamics are known, is to linearize the system about the operation point
% data-driven methods, such as ARX exists.

\textit{Lifted linear predictor} (LLP) is an artificial linear dynamical system designed to predict trajectories of a generally nonlinear dynamical system based on the current states (or measurements) and the inputs.
By \textit{lifting} we mean mapping the original state $x \in \mathbb{R}^n$ into a higher-dimensional space~$\mathbb{R}^N$, $N>n$, where the evolution of the state $x$ is approximately linear.
The idea of capturing nonlinear system's behavior linearly in lifted space dates back to the seminal works~\cite{koopman_hamiltonian_1931,koopman_dynamical_1932} of Koopman and von Neummann in the 1930s, with the resurgence of interest in mid 2000s with the works~\cite{mezic_comparison_2004,mezic_spectral_2005}.
Notably, the authors in~\cite{korda_linear_2018} introduced the use of LLP in conjunction with the Model Predictive Control (MPC).
Comprehensive picture of the applied Koopman Approach, discussing the theory and applications, can be found in~\cite{brunton_modern_2022}.

% ========= What is EDMD
There are several options to obtain the LLPs.
We focus on a data-driven approach, the \textit{Extended Dynamic Mode Decomposition} (EDMD) algorithm introduced in~\cite{williams_extending_2016}.
The EDMD presents an extremely simple method to obtain the control-oriented LLP from data, proving its efficiency in the control design for various systems~\cite{arbabi_data-driven_2018,korda_power_2018,do_controlled_2023}.
% Bounds on prediction errors for a type of LLP was theoretically derived in
% However, despite the method's efficiency, we show several practical problems that can arise in the process, failing to solve the particular task.
However, despite the method's efficiency, we show several practical problems that can arise in the process, failing to provide the desired control performance.

\subsection{Main Contribution}
In this paper, we address some less intuitive practical issues arising in identification of LLPs.
We present practical guidelines for identification and evaluation of linear predictors that address these issues.
As a result, better LLPs can be obtained.
We support the guidelines with two motivating examples as well as with references to relevant works.
The implementation of the examples are available at \href{https://github.com/aa4cc/Lifted-Linear-Predictors-Guidelines}{github.com/aa4cc/Lifted-Linear-Predictors-Guidelines}.

% In particular, we follow the approach developed in~\cite{korda_linear_2018}, where the Koopman operator approach was extended to controlled systems and a method for its approximation based on the Extended dynamic mode decomposition (EDMD) was proposed.
% ... how to practically obtain a sufficiently accurate predictor from given data.
% starting with some textbook examples but also showing more complex systems.

%=============================================================================
%=============================================================================
%=============================================================================

\section{Linear Predictors in Lifted Space}\label{sec:LLP}
% Our introduction mainly serves to fix the terminology and notation for the purpose of this work; if it is found too terse, we refer the reader to~\cite{brunton_modern_2022} or \cite{budisic_applied_2012} for more detailed introductions.
This section serves only as an overview -- providing definitions, fixing the notation and terminology for the purpose of this paper.
If it is found too terse, we refer the reader to~\cite{budisic_applied_2012,brunton_modern_2022} for more detailed description.

Consider a nonlinear discrete-time dynamical system
\begin{equation}\label{eq:nonlinear_sys}
  % \begin{split}
    x_{k+1} = f(x_k, u_k)\;, \\
    % y_k &= h(x_k) \;, 
  % \end{split}
\end{equation}
% $h(.)$ is a measurement function
% ${y \in \mathbb{R}^m}$,
where $f(\cdot)$ is a generally nonlinear transition mapping, and ${x \in \mathbb{R}^n}$, and ${u \in \mathbb{R}^p}$ are the system's state, output, and input, respectively.
The goal is to predict the trajectory of the system, given an initial condition $x_0$ at time $t_0$, and the sequence of the inputs $\{u_i\}_{i=0}^{N_\mathrm{p}}$, where $N_\mathrm{p}$ is the prediction horizon length.  
% (\blue{maybe use "+" notation})
% Let 
% \blue{...}
% At time $t_0$, the predictor is initialized with the current system's state $x_0$ and the sequence of the inputs $\{u_i\}_{i=0}^{N_\mathrm{p}}$, where $N_\mathrm{p}$ is the prediction horizon length.  
To this end, we define the \textit{lifted linear predictor} as a dynamical system
\begin{equation}\label{eq:lifted_predictor}
  \begin{split}
    z_0           &= \Psi(x_0) \;,\\
    z_{k+1}       &= Az_k + Bu_{k}\;,\\
    \hat{x}_{k} &= Cz_k \;, \\
  \end{split}
\end{equation}
where $z \in \mathbb{R}^N$ is the lifted state, $\Psi: \mathbb{R}^n \rightarrow \mathbb{R}^N$ is the lifting mapping, and $\hat{x}$ is the prediction of original system's state $x$.
The lifting mapping $\Psi(\cdot)$ is defined as
\begin{equation}
  \Psi(x) = \left[ \psi_1(x), \psi_2(x), \ldots, \psi_N(x)  \right]\tran \;,
\end{equation} 
where $\psi_i : \mathbb{R}^n \rightarrow \mathbb{R}$ is a scalar function called an \textit{observable}.
Apart from the LLP~\eqref{eq:lifted_predictor}, other forms of lifted predictors were introduced in literature that can have better prediction ability, see Remark~\ref{rem:different_predictors}.
However, we focus only on the LLP as it is immediately suited for linear control systems methods.

\begin{remark}\label{rem:different_predictors}
  The predictor can also be in the form of a \textit{bilinear} system
  \begin{equation}\label{eq:bilinear_predictor}
    \begin{split}
      z_{k+1} &= Az_k + (Bz_k)u_k \;, \\
      \hat{x}_{k} &= Cz_k \;,
    \end{split}
  \end{equation}
  which can to capture a more complex dynamical behaviors.
  Another form can be obtained by modifying the LLP (or the bilinear predictor) with the~\textit{project-and-lift method} proposed in~\cite{nuske_finite-data_2022}.
  In particular, for LLP~\eqref{eq:lifted_predictor}, the modified version is
  \begin{equation}\label{eq:project_and_lift_predictor}
    \begin{split}
      z_0         &= \Psi(x_0) \;,\\
      \tilde{z}_k &= \Psi(Cz_k) \;, \\
      z_{k+1}     &= A\tilde{z}_k + (Bz_k)u_k \;, \\
      \hat{x}_{k}   &= Cz_k \;.
    \end{split}
  \end{equation}
  Generally, the modified version~\eqref{eq:project_and_lift_predictor} should have better prediction ability on longer time horizons than the LLP~\cite{nuske_finite-data_2022}.
  However, both modifications~\eqref{eq:bilinear_predictor} and~\eqref{eq:project_and_lift_predictor} lose the linear structure.
\end{remark}

%==============================================================================================

\subsection{Koopman Operator for Autonomous Systems}\label{sec:koopman_operator}

To rigorously justify the construction of a linear predictor through state-space \textit{lifting}, we briefly describe the \textit{Koopman operator} approach for analysis of dynamical systems. 
% We start by introducing the Koopman operator for autonomous (uncontrolled) systems.
Consider a discrete-time dynamical system
\begin{equation}\label{eq:disc_non-linear_sys_Koopman}
        x^+ = f(x) \;,
\end{equation}
where $x, x^+ \in \mathcal{M}$ denote the current state, and the state in the next time step, respectively. Both are defined on some state space $\mathcal{M}$.
Instead of directly analyzing the generally nonlinear mapping $f$, we investigate how functions of the states, the \textit{observables}, evolve along the trajectories of the system.
Formally, an observable is a scalar function ${\psi:\mathcal{M} \rightarrow \mathbb{R}}$ that belongs to a typically infinite-dimensional space of functions $\mathcal{F}$.
The (discrete-time) \textit{Koopman operator} ${\mathcal{K}:\mathcal{F}\rightarrow\mathcal{F}}$ is then defined as 
\begin{equation}\label{eq:koopman_definition}
        ( \mathcal{K} \psi )(x) = \psi (f(x)) = \psi(x^+) \;.
\end{equation}
Thus, the Koopman operator $\mathcal{K}$ advances the observables ${\psi \in \mathcal{F}}$ from the current time step into the next one.

There are two key properties of the Koopman operator.
First, if the space of observables $\mathcal{F}$ contains the coordinate identity mappings $x\mapsto x(i)$ for all components $i$ of the state $x$, the operator fully captures the behavior of the nonlinear system~\eqref{eq:disc_non-linear_sys_Koopman}.
Second, it follows from the definition and the linearity of the addition in the function space that the operator is linear even if $\mathcal{T}$ is nonlinear.

Note, however, that the definition~\eqref{eq:koopman_definition} requires the space $\mathcal{F}$ to be invariant under the action of the operator $\mathcal{K}$.
For most systems, this leads to a requirement of the space $\mathcal{F}$ to be infinite dimensional -- containing infinitely many observables~$\psi{(x)}$ -- which precludes its use for control synthesis.

% Therefore, using the Koopman operator, the properties of a finite-dimensional nonlinear system can be fully described by a linear but infinite-dimensional operator.
% ======= NOTE: removed to save space =======
%  since for any two observables, $\psi_1$ and $\psi_2$, and scalar values $a_1$ and $a_2$, it holds
% \begin{equation}
    % \mathcal{K} \left(a_1\psi_1(x) + a_2\psi_2(x) \right)  =  a_1 (\mathcal{K}\psi_1)(x) + a_2 (\mathcal{K}\psi_2)(x) ,
% \end{equation}
% where we used the definition relation~\eqref{eq:koopman_definition} and the linearity of the addition in the function space.
% ===========================================
% While Koopman operator fully describes the properties of a nonlinear system, it is generally infinite-dimensional which precludes its use in control synthesis.
% Therefore, using the Koopman operator, the properties of a finite-dimensional nonlinear system can be fully described by a linear but infinite-dimensional operator.
% For control synthesis, the Koopman operator is not directly usable, as it is infinite dimensional.

There are two ways how the Koopman operator can be linked with the LLP~\eqref{eq:lifted_predictor}.
For some systems, one can find an invariant subspace of $\mathcal{F}$ spanned by a finite number of observables, while also containing the coordinate identity mappings.
The Koopman operator can then be represented by a finite-dimensional matrix, coinciding with the linear system's matrix.
% Such system can then be described in the form of LLP~\eqref{eq:lifted_predictor}.
Second, more generally applicable option, that we pursue in this paper, is to find a finite approximation to the Koopman operator via the algorithm described in Sec.~\ref{sec:edmd}.
% The main idea is to select observables, that approximate with sufficient accuracy the Koopman invariant subspace.

% Although the Koopman operator $\mathcal{K}$ completely describes the nonlinear system's behavior, the operator $\mathcal{K}$ is, for a general nonlinear system, infinite-dimensional which precludes its direct use in linear control schemes.

%==== Koopman Operator for Controlled Systems 
\subsection{Koopman Operator for Controlled Systems}\label{sec:koopman_operator_ctrl}
There are several ways of extending the Koopman operator to controlled systems.
For example, in~\cite{peitz_data-driven_2020}, the authors proposed treating a controlled system as an uncontrolled one with the input as system's parameter, see Remark~\ref{rem:interpolation_koopmans}.

We follow the extension proposed in~\cite{korda_linear_2018}, that associates a Koopman Operator with the dynamical system that evolves on the extended state space.
In particular, consider a controlled system
\begin{equation}\label{eq:disc_non-linear_sys_Koopman_ctrl}
        x^+ = f(x, u) \;.
\end{equation}
Let ${\ell (\mathcal{U})}$ be the space of all infinite input sequences ${u_\bullet = \{u_i\}_{i=0}^\infty}$, with $u_i \in \mathcal{U}$ and $\mathcal{U}$ being some input-space.
We denote the $i$th component of the sequence as $u_\bullet(i)$ and ${S}^*$ to be the (left) \textit{shift} operator, i.e., $\mathcal{S}^* u_{\bullet}(i) = u_{\bullet}(i+1)$.
By defining an augmented state $\chi = [x, u_\bullet]\tran$, we can rewrite the system~\eqref{eq:disc_non-linear_sys_Koopman_ctrl} as
\begin{equation}\label{eq:koopman_definition_ctrl}
        \chi^+ = f'(\chi) = 
        \begin{bmatrix}
                f \left( x, u_{\bullet}(0) \right) \\ \mathcal{S}^* u_\bullet 
        \end{bmatrix}
        \;.
\end{equation}
% and $u_{\bullet}[i]$ denotes $i$th element of the sequence $u_\bullet$. 
Let $\psi': \mathcal{M} \times \ell (\mathcal{U}) \rightarrow \mathbb{R}$ be an observable for the augmented state that belongs to a suitable space of functions $\mathcal{H}$. 
The Koopman operator $\mathcal{K}: \mathcal{H} \rightarrow \mathcal{H}$ for the controlled system~\eqref{eq:koopman_definition_ctrl} is then
\begin{equation}
        ( \mathcal{K} \psi' )(\chi) = \psi' (f'(\chi)) = \psi'(\chi^+) \;.
\end{equation}
Constructing a finite-dimensional \textit{approximation} to the operator $\mathcal{K}$ then induces a linear predictor in the form of the system~\eqref{eq:lifted_predictor}.

\begin{remark}\label{rem:interpolation_koopmans}
  Instead of a single Koopman operator describing the system's dynamics, a set of Koopman operators are identified.
  Each Koopman operator (parametrized by the input) then captures the system's behavior for a certain, constant, input.
  % In turn, the resulting LLP 
  % One can then interpolate between the different Koopman Operators.
  % When controlling the system with the MPC, this leads to mixed integer quadratic program -- a hybrid MPC.
\end{remark}

% \blue{We first describe what the linear predictor is}
%=============================================================================

\subsection{Extended Dynamic Mode Decomposition with Control}\label{sec:edmd}

The Extended Dynamic Mode Decomposition (EDMD) presents a simple way to approximate the Koopman operator from data, hence obtaining the LLP.
% identify the LLP from data.
The starting point is to select $N$ observables $\psi_1,\ldots,\psi_N$ and form the predictor's state $z$ from the measured state $x$ of the system~\eqref{eq:nonlinear_sys} as
\begin{equation}\label{eq:n_observables_general}
        z = \Psi(x) = [\psi_1(x), \psi_2(x), \ldots, \psi_n(x)]\tran\;.
\end{equation}
Next, we gather measured states $x_i$ and $x_i^+$ from the system and form a set of data 
\begin{equation}\label{eq:EDMD_matrices}
\begin{split}
    X           & = [x_1, x_2, \ldots, x_{N_\mr{d}}]\;, \\
    X_\mr{lift} & = [\Psi(x_1), \Psi(x_2), \ldots, \Psi(x_{N_\mr{d}})]\;, \\
    Y_\mr{lift} & = [\Psi(x_1^+), \Psi(x_2^+), \ldots, \Psi(x_{N_\mr{d}}^+)]\;, \\
    U           & = [u_1, u_2, \ldots, u_{N_\mr{d}}]\;,
\end{split}
\end{equation}
where $N_\mathrm{d}$ is the number of measurements, and the measurements satisfy a relation 
\begin{equation}\label{eq:1_one_step_output_relation}
        x_i^+ = f(x_i,u_i)   \;,      
\end{equation} 
that we wish to capture with the linear predictor. %and $\mathcal{T}(y_i,u_i)$ is an input-dependent transition mapping.
% \red{Add graphical depiction of the dataset}
Note that the states $x_i$ forming the set~\eqref{eq:EDMD_matrices} need not be temporally ordered, i.e., the states could be gathered from multiple trajectories.
The linear predictor~\eqref{eq:lifted_predictor} can be then identified by solving
% \begin{subequations}\label{eq:DMD_problem_def}
%         \begin{equation}
%                 \min_{A,B}      \| Y_\mr{lift} - AX_\mr{lift} - BU \|_\mr{F} \;,
%         \end{equation}
%         \begin{equation}\label{eq:DMD_problem_def_C_mat}
%                 \min_{C}        \| X - C X_\mr{lift} \|_\mr{F}  \;,
%         \end{equation}
% \end{subequations}
\begin{equation}\label{eq:DMD_problem_def} % Takes less space
  \begin{split}
    \min_{A,B}    &  \| Y_\mr{lift} - AX_\mr{lift} - BU \|_\mr{F} \;, \\
    \min_{C}      &  \| X - C X_\mr{lift} \|_\mr{F}  \;,
  \end{split}
\end{equation}
where $\| . \|_\mr{F}$ denotes a Frobenius norm of a matrix.
The solution to~\eqref{eq:DMD_problem_def} can be analytically obtained from
\begin{equation}
        \begin{bmatrix}
                A & B \\
                C & 0 
        \end{bmatrix}
        =
        \begin{bmatrix}
                Y_\mr{lift} \\
                X
        \end{bmatrix}        
        \begin{bmatrix}
                X_\mr{lift} \\
                U
        \end{bmatrix}\tran 
        \left(
        \begin{bmatrix}
                X_\mr{lift} \\
                U
        \end{bmatrix}
        \begin{bmatrix}
                X_\mr{lift} \\
                U
        \end{bmatrix}
        \tran 
        \right)^\dagger     \;,
\end{equation}
where $(.)^\dagger$ denotes the Moore--Penrose pseudoinverse of a matrix.
Note that when all components of the original state $x$ of the nonlinear system are incorporated in the set of observables, the matrix $C$ can be directly constructed by selecting the corresponding observables from~\eqref{eq:n_observables_general}.

% Even though the EDMD algorithm finds a representation (matrices $A,B$, and $C$) that minimizes \textit{one-step} , 
The convergence of the EDMD to the Koopman operator~\eqref{eq:koopman_definition} for autonomous systems was proven in~\cite{korda_convergence_2018}. 
This proof holds true when both the number of data points and the number of observables go to infinity.
However, from the practical point of view, one always works with finite number of both the data and the observables.
Additionally, identifying Koopman operator for controlled systems requires, generally, all possible input sequences, which is also unfeasible to obtain.

\subsection{Control Synthesis with LLP}

With identified LLP~\eqref{eq:lifted_predictor}, we can design a feedback controller with classical linear control systems method.
In this paper, we use two model-based techniques, a linear quadratic regulator (LQR) and model predictive control (MPC).

\subsubsection{LQR}
The discrete-time LQR minimizes the quadratic cost function
\begin{equation}\label{eq:lqr_cost_discrete}
  J(z,u) = \sum_{k = 0}^\infty \left(  z_k\tran Q z_k + u_k\tran R u_k + 2z_k\tran S u_k\right) \;,
\end{equation}
subjected to the system's dynamics~\eqref{eq:lifted_predictor}, and where $Q,R,S$ are cost matrices reflecting the desired control goal.
The control law minimizing the cost~\eqref{eq:lqr_cost_discrete} is 
\begin{equation}\label{eq:LQR_feedback_law}
  u_k = -Kz_k = K \Psi(x_k) \;,
\end{equation}
where the feedback gain $K$ can be computed from the solution to the discrete-time algebraic Riccati equation.
When the LQR is used for tracking, the control law changes to 
\begin{equation}
  u_k = -K\left(\Psi\left(z_k\right) - \Psi\left(z_\mathrm{ref}\right)\right) \;.
\end{equation}

\begin{remark}
  When using LLP within the LQR framework, the lifted states can serve not only for capturing the system's dynamics more accurately, but also can be directly penalized.
  The lifted states allow introducing penalization of arbitrary functions of state, extending the quadratic cost~\eqref{eq:lqr_cost_discrete}.  
  % For instance, one could be interested in minimizing the
  % For instance, one could be interested in minimizing the system's potential energy that depends nonlinearly on system's state. 
  %Thus, we could introduce an observable equal to the square of the velocity and penalize it in the performance index.
\end{remark}

\subsubsection{MPC}
The use of LLP within the MPC is coined as Koopman MPC (KMPC) in~\cite{korda_linear_2018}.
Here, we only state the main formulation and omit the details as they can be found in the referenced literature.
At every (discrete) time, the MPC solves the following optimization problem on the time horizon $N_\mathrm{p}$ 
\begin{equation}\label{eq:MPC_basic_opt_problem}
  \begin{split}
  \min_{u_k,z_k}                \quad &   J\left(\{z_k\}_{k=0}^{N_\mathrm{p}}  \;, \{ u_k \}_{k=0}^{N_\mathrm{p}} \right)   \;, \\
  \textnormal{subject to }      \quad &   z_{k+1} = Az_k + Bu_k,          \quad k=0,1,\dots,N_\mathrm{p}-1   \;, \\
                                \quad &   e_k = r_k - Cz_k                                        \;, \\
                                \quad &   z_{\rm min} \leq  z_k \leq z_{\rm max}                  \;, \\
                                \quad &   u_{\rm min} \leq  u_k \leq u_{\rm max}                  \;, \\
  \textnormal{parameters}       \quad &   z_0 = \Psi(x_t)                                         \;, \\
                                \quad &   r_k = \textnormal{given}  \;,  \quad k =0,1,\dots,N_\mathrm{p}    \;, 
  \end{split}
\end{equation}
where $u_{\rm min/max}$ are bounds on inputs, $z_{\rm min/max}$ are bound on lifted states, and $r_k$ is the reference signal.
Only the first element of the optimal sequence $u_0$ is used, and the optimization problem is solved again in the next time step. 
The cost function $J$ is
\begin{equation}\label{eq:mpc_cost_discrete}
  J(z,u) = e_{N_\mathrm{p}}\tran Q_N e_{N_\mathrm{p}} + \sum_{k = 0}^{N_\mathrm{p} - 1} \left(  e_k\tran Q e_k + u_k\tran R u_k\right) \;.
\end{equation}

%=============================================================================\
%=============================================================================
%=============================================================================

\section{Practical Guidelines}\label{sec:Practical_Guidelines}

This section presents some key and less intuitive practical guidelines for identification of LLPs using the EDMD algorithm.
We support the guidelines with several motivating examples in Sec.~\ref{sec:examples}.

%=============================================================================

\subsection{Select Appropriate Metric to Assess Predictors Effectively}\label{sec:evaluation_def}
As for any data-driven method, it is important to evaluate the accuracy of the obtained model.
In the following subsection, we define and discuss different metrics for assessing the performance of LLPs.
% \blue{To-be section's content:
% The type of the metric is important.
% Some metrics (errors) are suitable for different use-cases.
% }
% The EDMD algorithm, while it finds the matrices $A,B$, and $C$  
% ... to detect overfitting.
% We define several options. to evaluate the accuracy of the identified LLP .
Let $D = (x_i, x_i^+, u_i)$, $i = 1, \ldots N_\mathrm{d}$ be a dataset of samples from evaluation trajectories, and let the LLP be given by the matrix triplet $A,B,C$ and the lifting function $\Psi$.
We argue that the type of error selected for evaluation, depends on the purpose of the LLP.

Probably the most natural option would be to evaluate the error
\begin{equation}\label{eq:e_unlift}
  \epsilon_\mathrm{projected} = \frac{1}{N_\mathrm{d}}\sum_{i = 1}^{N_\mathrm{d}} \lVert  x_i^+ - C( A \Psi(x_i) + B u_i  )  \rVert \;,
\end{equation}
where $\lVert.\rVert$ is the 2-norm.
The error~\eqref{eq:e_unlift} reflects how the LLP predicts time evolution of the system's original states, excluding the lifted states.
However, note that the error~\eqref{eq:e_unlift} is not directly minimized by the EDMD algorithm (see Eq.~\eqref{eq:DMD_problem_def}).
In particular, the EDMD finds the LLP that minimizes the error in prediction of the \textit{full lifted state}, that is
\begin{equation}\label{eq:error_lift_state_one_step}
  \epsilon_\mathrm{lifted} = \frac{1}{N_\mathrm{d}}\sum_{i = 1}^{N_\mathrm{d}} \lVert  \Psi(x_i^+) - ( A \Psi(x_i) + B u_i  )  \rVert \;.
\end{equation}
Therefore, while adding more identification data or more observables should theoretically result in lower $\epsilon_\mathrm{lifted}$, the error $\epsilon_\mathrm{projected}$ reflecting the prediction accuracy of relevant variables might be worse as the EDMD could prioritize capturing the lifted states.
In other words, the EDMD does not explicitly guarantee minimization of the error~\eqref{eq:e_unlift} so also the error~\eqref{eq:error_lift_state_one_step} should be taken into account in the identification process.

Lastly, for some control methods, especially the MPC, more important is to evaluate the LLP's ability to predict systems' trajectory on a certain time horizon with a length $N_\mathrm{p}$.
% We define a \textit{prediction error} on a prediction horizon 
Consider a trajectory with an initial state $x_0$, the input sequence $\{u_i\}_0^{N_\mathrm{p}-1}$, and corresponding output sequence $\{x_k\}_0^{N_\mathrm{p}}$ satisfying the relation~\eqref{eq:nonlinear_sys}.
We define the prediction error as
\begin{equation}\label{eq:prediction_horizon_error}
  \epsilon_\mathrm{prediction} = \frac{1}{N_\mathrm{d}}\sum_{i = 0}^{N_\mathrm{d}} \sum_{k = 0}^{N_\mathrm{p}-1} \lVert x_{k+1} - C( A \Psi(x_k) + B u_k  )  \rVert^2 \;.
\end{equation}
Moreover, the metric can be used to select the length of the prediction horizon $N_\mathrm{p}$ of the MPC.

% In particular, 
% \blue{the prediction horizon $N_\mathrm{p}$ for MPC can be selected in order to use a LLP that}
% \blue{Compute prediction error on the same prediction horizon used in MPC or vice-versa: choose prediction horizon based on the accuracy of prediction.}

\begin{remark}
  An important step in assessing LLP's accuracy is taking into account different scales of state variables.
  % .. so the
  % and there are many options how the state can be normalized.
  For example, all states can be scaled into an interval $\left\langle -1, 1 \right\rangle$ or the errors themselves otherwise normalized.
\end{remark}

% \blue{Maybe} The lifted state error might be a good indicate of right choice of lifting functions.
% There are several options to evaluate the accuracy of the identified LLP
% With the identified LLP~\eqref{eq:lifted_predictor}, given by the matrix triplet $A,B,C$, and the lifting function $\Psi$, there are several options to evaluate its accuracy to capture the original system's dynamics.
% We gather an additional dataset, consisting of triplets 
% Naturally, the dataset $D$ should be sampled 
% We define several types of errors.
% The second type of error is the 
% However, 
% One could argue that since the observables are 
% Another way to evaluate the LLP is to consider only the prediction accuracy for the original system's state, as we can regard the additional observables only as \textit{auxiliary} variables.
% We define the 
% Examining the error~\eqref{eq:error_lift_state_one_step} is 

%=============================================================================

% \subsection{Carefully Choose Lifting Functions; Sometimes, Less is More}
\subsection{Carefully Choose Lifting Functions}

% \blue{To-be section's content:
% Selection of lifting function is crucial.
% Sometimes, more is less.
% There are many options, working for different systems.
% }
% The selection of lifting function is the central challenge in the EDMD algorithm.

Following the EDMD algorithm, the first and the most challenging step in the identification process is selection of the observables $\psi_i$.
As explained in Sec.~\ref{sec:koopman_operator}, the aim is at approximating the Koopman operator on some finite-dimensional function space. Typically, a basis of such space is selected.

In~\cite{mamakoukas_local_2019}, the authors proposed using higher order derivatives of the underlying nonlinear dynamics, motivated by choice of lifting functions for polynomial systems.
Delay embeddings are another option, allowing to reconstruct the system's state from the system's output, which can be useful when only sparse measurements are available~\cite{arbabi_data-driven_2018}.
% A different approach is to consider the 
The set of observables could also result from optimization problem, as proposed in~\cite{korda_optimal_2020}, exploiting the system dynamics and the state-space geometry.

Generally, we should aim to choose observables that accurately represent the problem's dynamics, focusing on those that capture well the manifold on which the system evolves. 
Prioritizing few specific problem-related observables over a general function space basis can often yield superior results -- \textit{sometimes, less is more}.

Another approach presented, e.g., in~\cite{lusch_deep_2018} is using a neural network as a lifting function.
This approach is based on learning the eigenfunctions of the Koopman operator which then yield the LLP.

% distinct from  the EDMD algorithm, involves utilizing a neural network as a lifting function, as presented . 
% It corresponds to learning the eigenfunctions of the Koopman operator.
% \blue{
% There are many options:
% \begin{itemize}
  % \item Delay embeddings
  % \item Polynomial basis
  % \item Radial basis functions (RBF)
  % \item Right-hand side of nonlinear functions to be identified
% \end{itemize}
% }

% \blue{Overfitting}

%=============================================================================

%\subsection{Control-oriented Data Sampling}\label{sec:guideline_quality_of_data}
% \subsection{Quality of Data}% \subsection{Select Representative Data}
\subsection{Identify LLPs with Data Similar to Intended Closed-Loop Trajectories}

% \blue{To-be section's content:
% Using Koopman-based predictor for control is problematic, because LLP (as it is linear system) has only one equilibrium or infinitely many.
% Nonlinear system can have several equilibria: it is unfeasible to obtain LLP that correctly captures the system's behavior with more equilibria.
% \textbf{guideline:} Select trajectories, that are close to the intended closed-loop trajectories.
% }

The next step in identification is collecting the data.
Generally, it is not feasible, nor desired to sample the data from the whole state space.
In~\cite{brunton_modern_2022}, authors illustrate the importance of the identification data for the \textit{Duffing Oscillator}, a system with three equilibria.
Based on the Koopman operator analysis of the system for controlled systems, it is important to choose representative both the control sequences and the system's trajectories.
Thus, the data should be ideally sampled from the trajectories that represent the final closed-loop trajectories.
Therefore, it is advantageous to identify an input that closely approximates the final closed-loop input and subsequently -- as he samples should be independent and identically distributed (i.i.d.) -- utilize a randomized signal that closely resembles it.
Note that using only closed-loop data could cause problems, as the feedback creates correlation between unmeasurable noise (in system's output) and the system's input, see~\cite{forssell_closed-loop_1999} for details.

% which creates correlation between the samples, 
% Also, the closed-loop identification is problematic  
% We showcase this in the Sec.~\ref{sec:example_pendulum}, showing control of an inverted pendulum.
% the example of system with two equilibria was shown.
% Overfitting on given trajectories.
% The correct selection of training data must be especially considered when 
% Similar 

%=============================================================================

\subsection{Inspect Identified Predictors}

While the LLP can be perceived only as a \textit{black box}, predicting system dynamics with adequate accuracy, its quality can be substantially enhanced through careful inspection and subsequent adjustment of the identification process.
For example, as the EDMD algorithm is purely data-driven, the identification can result in a LLP that allows non-physical coupling between the states.
Also, the \textit{stability} of the system (or system's equilibria) might not be preserved.
Finally, when the number of selected observables is significantly higher than the original system's order ($N \gg n$), a low-dimensional approximation to the LLP can be created.
Such an approximation may be necessary for certain control synthesis methods.
Details on order reduction methods in context of EDMD can be found, e.g., in~\cite{brunton_data-driven_2022}. 

% Finally, if the order of the resulted LLP is high, it might be beneficial (e.g., for control synthesis) to obtain a low-dimensional approximation to the LLP.
% A ... insight into the 

% the if number of observables is high, while a low-dimensional  the resulted high-order LLP can approximated by a low-rank  

% We showcase this in Sec.~\ref{sec:LLP_ID_example_two_wheel_robot}, where the EDMD algorithm leads to a LLP that allows non-physical coupling between the states.

% NOTE: add something about eigenvalues of the matrix A.

%=============================================================================
%=============================================================================
%=============================================================================

\section{Motivated Examples of LLP Identification and Application for Control}\label{sec:examples}

In this section, we apply the method described in Sec.~\ref{sec:LLP} to obtain LLP for several systems.
% The motivation of this is to highlight caveats in the identification process.
% Therefore, we intentionally present wrong choices and ways to solve them.
% Therefore, we first present a working solution and then several options that 
The whole method consists of acquiring data, selection of observables, evaluation of the LLP, and lastly, control design with control evaluation.
For the sake of brevity, we omit some details in control design as the main focus is on LLP identification, rather than the control synthesis.
The details can be found in the provided code.

In all examples, the systems are numerically integrated using the 4th order Runge-Kutta algorithm with an integration step ${T_\mathrm{s} = \SI{0.01}{\second}}$.
We first present a correct solution, and then we introduce some pathologies into the process to highlight the importance of some aspects, motivating our guidelines.
Note that our aim is not to present the best solution to the particular problem.
Rather, the examples serve to show the intricacies in LLP identification.

% =====================================

\subsection{Swing-up of Inverted Pendulum}\label{sec:example_pendulum}
Consider a damped pendulum, driven by a torque $u$, and with the system's origin $x_\mathrm{e}$ shifted into the unstable equilibrium -- the upward position.
Let $\varphi$ and $\omega$ be pendulum's angle and the angular velocity, respectively.
The system's model is 
\begin{equation}\label{eq:mathematical_pendulum}
  \dot{x}
  =
  \begin{bmatrix}
    \dot{\varphi} \\
    \dot{\omega}
  \end{bmatrix}
  =
  \begin{bmatrix}
    \omega \\
    \frac{g}{l}\sin(\varphi) - \frac{\gamma}{ml^2}\omega + \frac{1}{ml^2}u 
  \end{bmatrix}\;,
\end{equation}
where $g$ is the gravity, $\gamma$ is the damping coefficient, and $m$, $l$ are the pendulum's mass and length, respectively.
Both states, $\varphi$ and $\omega$, are measured.

We aim to design controller that achieves the \textit{swing-up} task, i.e., driving the system from a state ${x = [\pi, 0]\tran}$ (the downward position) into the state ${x_\mathrm{e} =[0, 0]\tran}$ (the upward position).
Since we do not consider limits on the input $u$, the task can be simply achieved by LQR with local linearization at $x_\mathrm{e}$.
We wish to identify LLP to design a controller with better performance.

We gather the identification data from closed-loop trajectories, using a proportional control law $u = -k_\mathrm{p}\varphi$.
To obtain rich dataset, each trajectory is initialized in a random state while also the gain $k_\mathrm{p}$ is randomized.
Additionally, the closed-loop input is perturbed by a random signal.
In total, we collect 2100 trajectories from which we use 100 trajectories for evaluation only.
Each trajectory was $\SI{0.5}{\second}$ in length.
% in the lifting function, we  
As the observables, we select the original states together with single additional function, so the lifting function is
\begin{equation}
  \Psi(x) = [\varphi, \omega, \sin{(\varphi)}]\tran \;.
\end{equation}
This is an example of problem-specific lifting, motivated by the sine function's presence in system's dynamics~\eqref{eq:mathematical_pendulum}.

\begin{figure}
    \centering
    \includegraphics[width=1\linewidth]{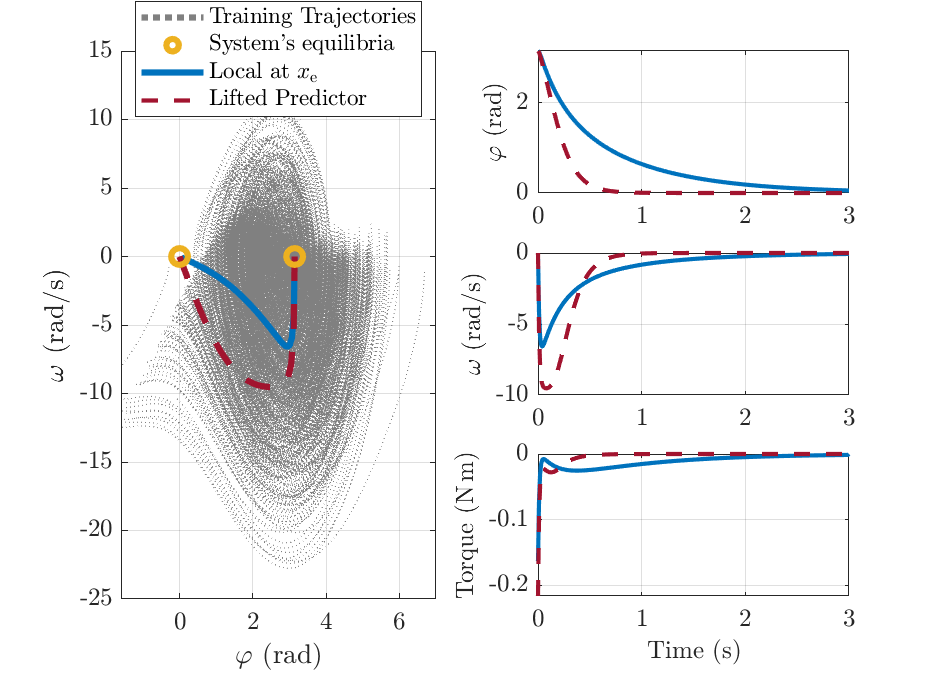}
    \caption{The figure's right part displays the phase portrait of training trajectories for the swing-up task, together with comparison of closed-loop performances. On the left are time series.}\label{fig:pendulum_working}
\end{figure}

Then, we devise an LQR controller using the LLP, and for comparison, another LQR controller based on local linearization around the system's origin, referred to as a \textit{local predictor}.
The Fig.~\ref{fig:pendulum_working} shows a phase-portrait with training trajectories together with LQR closed-loop trajectories using two linear models.
We can see that the lifted predictor exhibits superior performance, even when the penalty matrices $Q$ and $R$ of both LQRs are the same.

% =====================================

\subsubsection{Pathology \#1: Collecting Identification Data}

\begin{figure}
  \centering
  \includegraphics[width=1\linewidth]{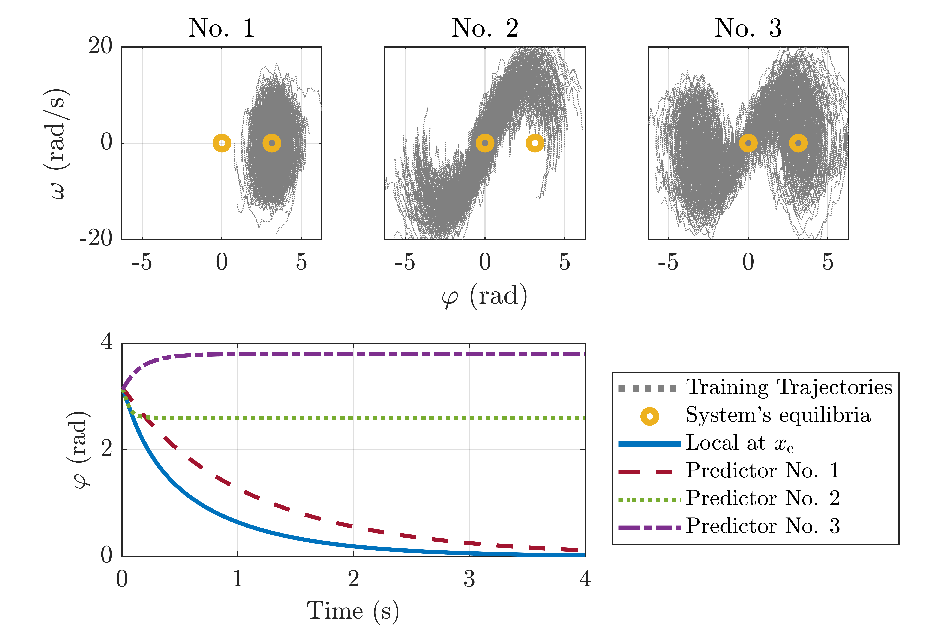}
  \caption{ Three examples of non-representative identification data for the swing-up task. 
            In the figure's top part are the phase portraits of identification data, at the bottom are comparisons of closed-loop performances}\label{fig:pendulum_pathology_data}
\end{figure}

To collect identification data, an alternative option to the closed-loop input might be using (pseudo) random signal. 
We also present three different choices of initial states for the data: initialization near the stable equilibrium, near the unstable equilibrium, and uniformly distributed in the interval ${\varphi_0 = (-\pi, \pi)}$.
The phase portraits of the identification data are displayed in Fig.~\ref{fig:pendulum_pathology_data}.
In the bottom part of the figure, we then compare the closed-loop control.
We can see, that only the first option of sampling results in a successful swing-up (albeit worse than local linearization), while the other two options fail.
The main challenge of this task is that the swing-up-achieving trajectory contains two equilibria. This is a problem, as a linear system cannot have two or more isolated equilibria.

While sampling the portion of the state space in an open loop may lead to a satisfactory behavior, the perturbed closed-loop data present more straightforward option.

\subsubsection{Pathology \#2: Choice of Lifting Functions}
In the spirit of the Koopman operator, we can try to obtain better representation of the system's dynamics by introducing more lifting function, hence to better approximate the Koopman operator.
We consider three additional options, differing in the basis functions: the polynomial basis, \textit{thin plate splines} radial basis functions (TPS-RBF), and \textit{Gaussian} RBF.
For each option, the lifting function $\Psi(x)$ consists of the original states $\varphi$ and $\omega$, and then additional 100 observables
\begin{equation}
  \begin{split}
    \psi_i^\mathrm{TPS} &= \lVert \varphi - a_i \rVert^2 \log \lVert \varphi - a_i \rVert  \;, \\
    \psi_i^\mathrm{pol} &= \varphi^{(i-1)} \;,  \\
    \psi_i^\mathrm{gauss} &= \exp \left( - (\varphi - b_i)^2/(2c_i^2) \right) \;, 
  \end{split}
\end{equation}
where the parameters $a_i, b_i, c_i$,  ${i = 1, \ldots, 100}$ were sampled randomly.
% Moreover, we also consider a system extended with the following singular observable:
% \begin{equation}
%   \psi^\mathrm{sine} = \sin{(\varphi)}.
% \end{equation}
% This is an example of problem-specific lifting, motivated by the sine function's presence in the governing equation's right-hand side.

\begin{figure}
  \centering
  \includegraphics[width=1\linewidth]{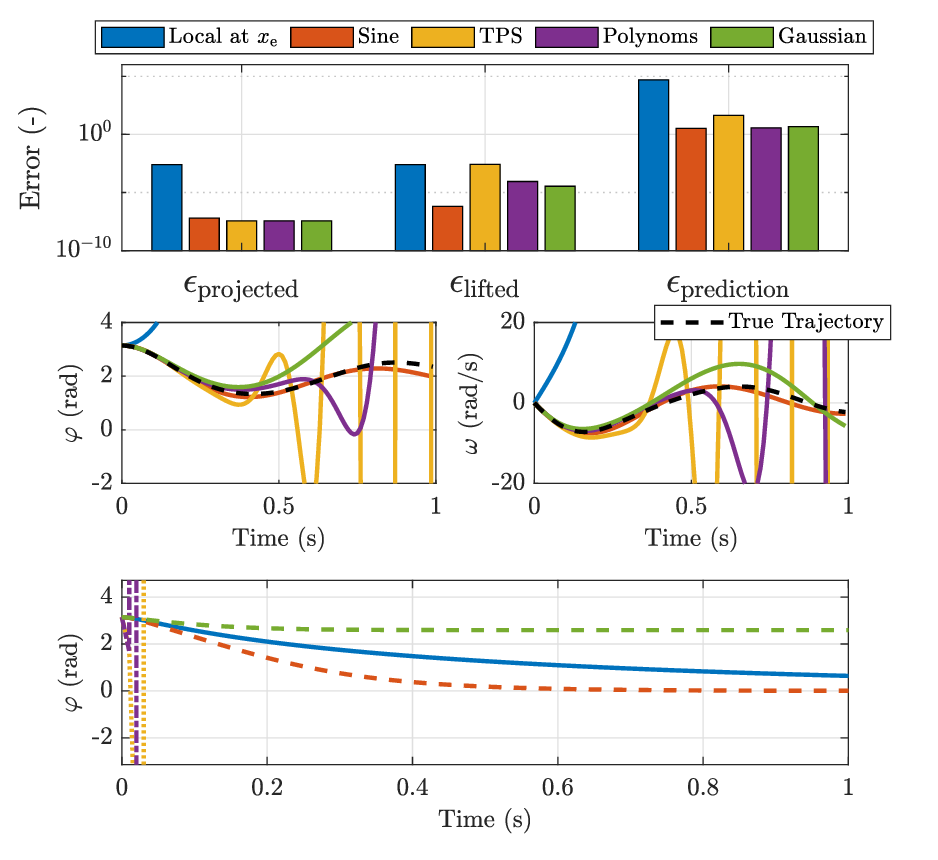}
  \caption{Comparison of different lifting functions. The figure, from top to bottom, shows the comparison of evaluated errors (in the logarithmic scale), prediction on a finite horizon, and the time series of the closed-loop control}\label{fig:pendulum_pathology_lifting}
\end{figure}

The comparison of the predictors in shown in Fig.~\ref{fig:pendulum_pathology_lifting}.
To assess the accuracy of the predictors, we first compare (in logarithmic scale) all the metrics introduced in Sec.~\ref{sec:evaluation_def}.
We also show a comparison of predictors' ability to capture a single trajectory, given its initial state and the sequence of inputs.
Lastly, we show the closed-loop performance.

An interesting observation is that while the projected errors $\epsilon_\mathrm{projected}$ of the predictors with many observables are better than the error of the \textit{local} or the \textit{sine} predictors, the resulting control is significantly worse.
We found that when using LLP for designing LQR, the most relevant metric appears to be the $\epsilon_\mathrm{lifted}$ since the feedback law~\eqref{eq:LQR_feedback_law} generally uses values of all lifted states $z$, even though the lifted states are not directly penalized.

% Good measure for the resulting control performance appears to be the prediction error. 
% It is the lowest for the sine lifting, which also performs the best in the closed-loop setting.

% Let us first identify 
% We deal with the problem by first identifying a LLP that we then use to design a linear-quadratic regulator (LQR), achieving the goal.
% \subsection{Correct Solution}
% % The main 
% As the first step, we gather the data.
% \subsubsection{}
% We compare three options to gather the identification data. 
% To gather the identification data, we 

% We tackle the problem by designing a linear-quadratic regulator (LQR) using the LLP.
% We describe two options how to gather identification data for the EDMD algorithm.
% For comparison, we also design a LQR based on the linear approximation of the system~\eqref{eq:mathematical_pendulum}.
%\paragraph{LLP Identification}
% \subsection{Forced Duffing Oscillator}
% \subsection{Fluid Flow}

% ===================================

\subsection{Two-wheeled Inverted Pendulum Balancing Robot}\label{sec:LLP_ID_example_two_wheel_robot}

\begin{figure}
    \centering
    \includegraphics[width=1\linewidth]{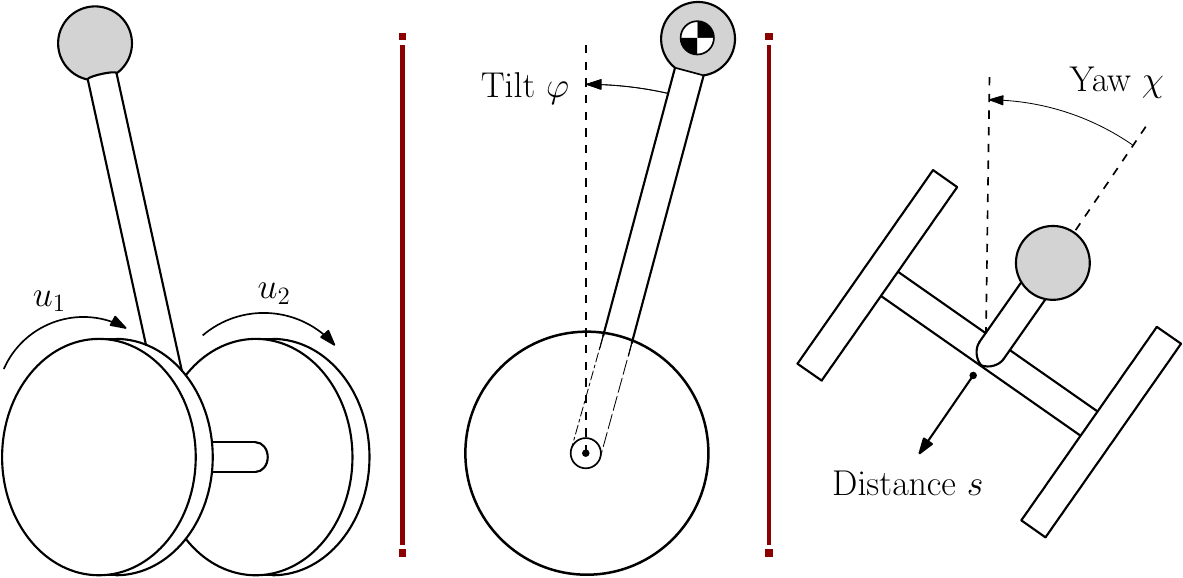}
    \caption{Wheeled Balancing Robot}\label{fig:temp_wheeled_robot_diag}
\end{figure}

We consider a system depicted in Fig.~\ref{fig:temp_wheeled_robot_diag}, a two-wheeled balancing robot moving on a flat surface with three degrees of freedom: the traveled distance $s$, tilt angle $\varphi$, and yaw angle $\chi$.
Let $q = [s,\varphi, \chi]\tran$ be the vector of generalized coordinates, and $u = [u_1, u_2]\tran$ be the system's input -- torques applied on the wheels.
The robot's dynamics (see~\cite{kim_dynamic_2015} for full derivations) can be described by the equation
\begin{equation}\label{eq:ex_sys_wheeled_robot}
  \begin{split}
    \dot{q} &= M(q)^{-1} \left(  Bu - C(q, \dot{q})\dot{q} - D\dot{q} - G(q)  \right) \;,
  \end{split}
\end{equation}
where $M(q)$, $C(q, \dot{q})$, $D$, $G(q)$, $B$ are inertia, Coriolis, dissipation, gravity, and input matrices, respectively.
The goal is maintaining the robot in the upward position while allowing to track a reference on the travel speed $\dot{s}_\mathrm{ref}$ and the yaw rate $\dot{\chi}_\mathrm{ref}$.
To address the task, we identify the LLP and use the predictor within the MPC.

We gather the identification and verification data from closed-loop trajectories.
To this end, we designed a LQR controller stabilizing the system to the equilibrium in the origin.
The LQR design was based on a local linearization of the system~\eqref{eq:ex_sys_wheeled_robot} about the system's origin.
In total, we gathered 2000 trajectories, where the half was used for identification, and the rest for validation.
Each trajectory was initialized with randomized coordinates $q$.
We selected the references $s_\mathrm{ref}$ and $\chi_\mathrm{ref}$ to be sinusoidal with randomized amplitude.

Let the system's state be $x = [\dot{q}, q]\tran$.
We select the lifting function for LLP as
\begin{equation}\label{eq:robot_lifting_correct}
  \Psi(x) = [\dot{s}, \dot{\varphi}, \dot{\chi}, \varphi, \sin(\varphi)]\tran\;.
\end{equation}
That is, we omit the states $s$ and $\chi$ as they should not, by the assumption, affect the system's dynamics, i.e., the dynamics should be invariant w.r.t the travelled distance $s$ and the yaw $\chi$.
Also, if needed, both states can be numerically recovered from their derivatives.
We add the observable $\psi(x) = \sin(\varphi)$ to capture the pendulum-like motion of the robot.

% and also with sinusoidal reference with  trajectories on the .
% and controlled with an input $u = -Kx$, where $K$ is the feedback gain.
% that uses a local linearization of the system~\eqref{eq:ex_sys_wheeled_robot} about the system origin.

\begin{figure}
  \centering
  \includegraphics[width=1\linewidth]{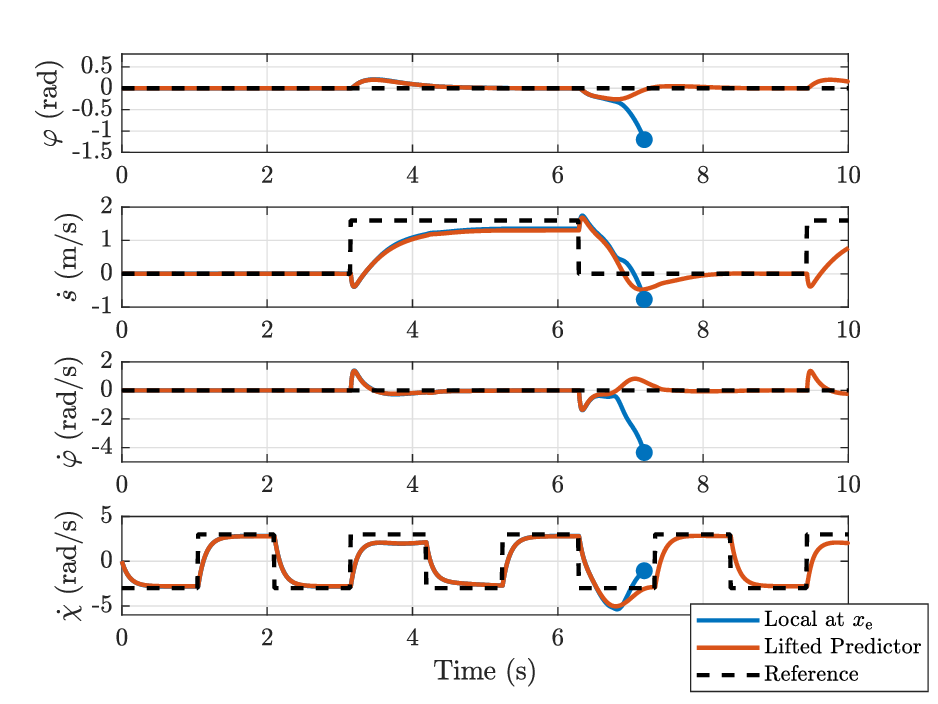}
  \caption{Comparison of closed-loop performances of MPC based on a local predictor or on the identified LLP. The MPC with local predictor fails to stabilize the system around $t=\SI{7}{\second}$}\label{fig:robot_KMPC_Local_comparison}
\end{figure}	

We designed two MPC controllers, one based on local linearization, and one based on identified LLP.
The Fig.~\ref{fig:robot_KMPC_Local_comparison} shows a comparison of the closed-loop performances.
The MPC with LLP exhibits better performance as the Local linearization fails to stabilize the system.
Even with only one extra lifting function, the prediction error~\eqref{eq:prediction_horizon_error} of the LLP was better by $\approx\SI{20}{\percent}$ than the local linearization.

% Therefore, for the defined goal, we can extract  define a reduced state $x = [\varphi, \dot{s}, \dot{\varphi}, \dot{\psi}]\tran$ 
% We define  as the system's state and assume that all components are measured.
% By observing the system's dynamics, 
% As we assume that the robot is moving on a level plane
% Since the dynamics of the system is invari

% ===================================

\subsubsection{Pathology \#1: Non-physical State Coupling}

\begin{figure}
  \centering
  \includegraphics[width=1\linewidth]{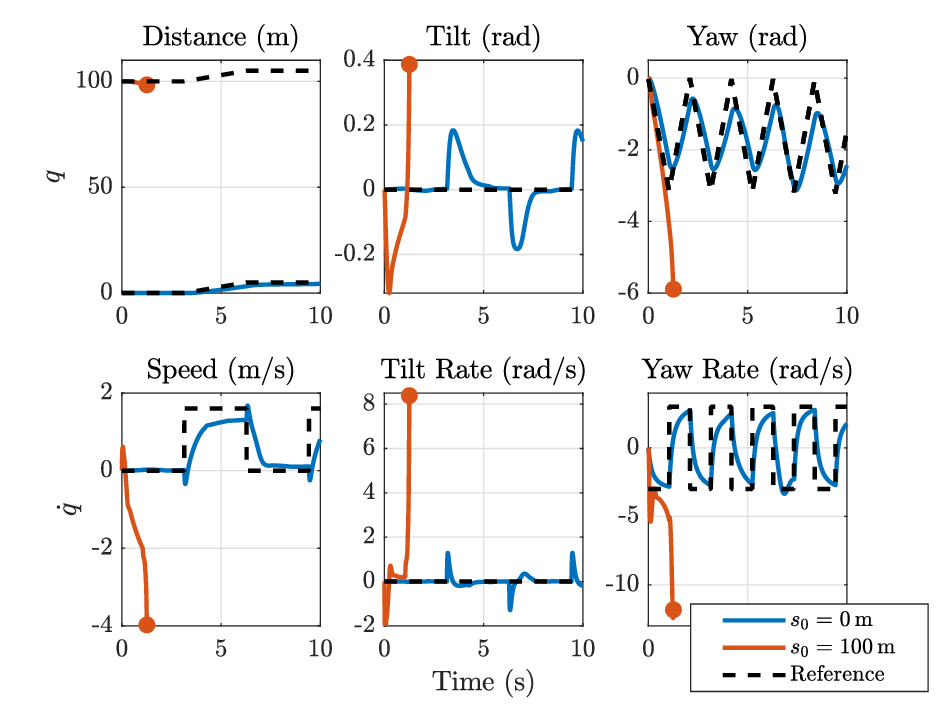}
  \caption{Demonstration of how non-physical coupling between the states affect the control performance. Shown are two trajectories, initialized in different distance $s_0$}\label{fig:robot_KMPC_pathology_coupling}
\end{figure}

Let us now identify a LLP for the system~\eqref{eq:ex_sys_wheeled_robot} with a lifting function that includes all the robot's states
\begin{equation}
  \Psi(x) = [\dot{s}, \dot{\varphi}, \dot{\chi}, s, \varphi, \chi, \sin\varphi]\tran\;.
\end{equation}
We repeat the whole process.
The Fig.~\ref{fig:robot_KMPC_pathology_coupling} shows the closed-loop performance of KMPC for the system initialized in two different initial states, differing in the initial distance $s_0$.

\begin{figure}
  \centering
  \includegraphics[width=1\linewidth]{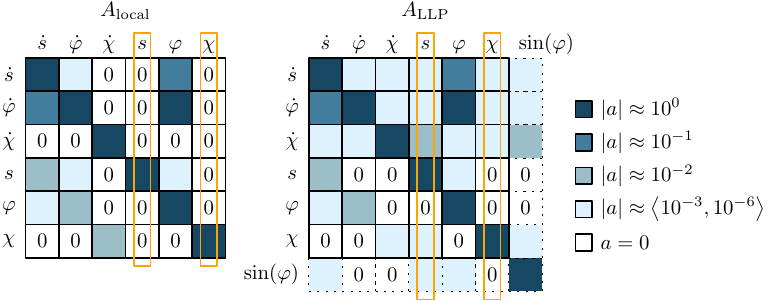}
  \caption{\textit{Heatmaps} of the system's matrices $A_\mathrm{local}$ and $A_\mathrm{LLP}$ showing non-physical coupling between the states for lifted predictor}\label{fig:heat_map_v2}
\end{figure}

When the system is initialized with $s_0 = 0$, the performance is comparable to the previous case with the lifting function~\eqref{eq:robot_lifting_correct}.
However, when the system is initialized with $s_0 = 100$, which should not influence the system, the controller fails to stabilize the system.
To gain an insight into this behavior, let us examine the identified system matrix $A_\mathrm{LLP}$ of the LLP and compare it with a system matrix $A_\mathrm{local}$ obtained from local predictor, see Fig.~\ref{fig:heat_map_v2}.

As evident from the $A_\mathrm{LLP}$ matrix, there exists significant coupling between $s$, $\chi$, and the remaining states. As previously noted, the system's dynamics are expected to remain invariant with respect to both the traveled distance $s$ and yaw rate $\chi$. The LLP resulting from the EDMD is not imposing such constraints on the resulting model.

% However, when we examine the identified system's matrix $A$, for example, the prediction of the state $\dot{\varph
% Let $q = $ be a vector of generalized coordinates, where $s$ is robot traveled distance, $\phi$ ti
% there should not be a coupling between the states $s$ and $\psi$,and the states $\varphi$ and $\omega$,
% \begin{equation}
%   \begin{split}
%     % {\varphi}_{k+1} &= 0.014s  \\
%     \dot{\varphi}_{k+1}^\mathrm{local} &= g(z_k) + 0.004s  \\
%     \dot{\varphi}_{k+1}^\mathrm{LLP} & =  [0.14, 0.99, 0, 0.14, 0.99, 0]x
%   \end{split} 
% \end{equation} 
% where $g(z_k)$ are other terms.
% However, 

% EDMD is not taking  to this physical interpretation.
% Using the assumption that the robot is moving on a flat surface, we can observe that the system's dynamics should be invariant w.r.t the distance $s$ and yaw angle $\psi$.
% That is, the absolute values of the states $s$ and $\psi$ should not influence other states.
% While the prediction accuracy of the LLP is superior to the accuracy of the linear approximation, the non-physical coupling 
% In this example, the problem can be easily fixed by removing the measurements of the states $s$ and $\psi$ from the identification
% Thus using only the states $ $ for identification.
% If the states are required for control, they can be added into the system manually.

%=============================================================================
%=============================================================================
%=============================================================================

\section{Conclusion}\label{sec::Conclusion}

This paper introduced practical guidelines for utilizing EDMD in the identification of control-oriented lifted linear predictors. 
We have demonstrated a range of challenges that may arise when employing EDMD for LLP identification, along with strategies to address them. 
Our future efforts will concentrate on expanding this set of guidelines to establish a comprehensive methodology for LLP identification.

\bibliography{biblio}

\begin{thebibliography}{10}

\bibitem{koopman_hamiltonian_1931}
Koopman BO.
\newblock Hamiltonian {Systems} and {Transformation} in {Hilbert} {Space}.
\newblock Proceedings of the National Academy of Sciences. 1931
  May;17(5):315--318.
\newblock Publisher: Proceedings of the National Academy of Sciences.

\bibitem{koopman_dynamical_1932}
Koopman BO, Neumann Jv.
\newblock Dynamical {Systems} of {Continuous} {Spectra}.
\newblock Proceedings of the National Academy of Sciences. 1932
  Mar;18(3):255--263.
\newblock Publisher: Proceedings of the National Academy of Sciences.

\bibitem{mezic_comparison_2004}
Mezić I, Banaszuk A.
\newblock Comparison of systems with complex behavior.
\newblock Physica D: Nonlinear Phenomena. 2004 Oct;197(1):101--133.

\bibitem{mezic_spectral_2005}
Mezić I.
\newblock Spectral {Properties} of {Dynamical} {Systems}, {Model} {Reduction}
  and {Decompositions}.
\newblock Nonlinear Dynamics. 2005 Aug;41(1):309--325.

\bibitem{korda_linear_2018}
Korda M, Mezić I.
\newblock Linear predictors for nonlinear dynamical systems: {Koopman} operator
  meets model predictive control.
\newblock Automatica. 2018 Jul;93:149--160.

\bibitem{brunton_modern_2022}
Brunton SL, Budišić M, Kaiser E, Kutz JN.
\newblock Modern {Koopman} {Theory} for {Dynamical} {Systems}.
\newblock SIAM Review. 2022 May;64(2):229--340.
\newblock Publisher: Society for Industrial and Applied Mathematics.

\bibitem{williams_extending_2016}
Williams MO, Hemati MS, Dawson STM, Kevrekidis IG, Rowley CW.
\newblock Extending {Data}-{Driven} {Koopman} {Analysis} to {Actuated}
  {Systems}.
\newblock IFAC-PapersOnLine. 2016 Jan;49(18):704--709.

\bibitem{arbabi_data-driven_2018}
Arbabi H, Korda M, Mezić I.
\newblock A {Data}-{Driven} {Koopman} {Model} {Predictive} {Control}
  {Framework} for {Nonlinear} {Partial} {Differential} {Equations}.
\newblock In: 2018 {IEEE} {Conference} on {Decision} and {Control} ({CDC});
  2018. p. 6409--6414.

\bibitem{korda_power_2018}
Korda M, Susuki Y, Mezić I.
\newblock Power grid transient stabilization using {Koopman} model predictive
  control.
\newblock IFAC-PapersOnLine. 2018 Jan;51(28):297--302.

\bibitem{do_controlled_2023}
Do L, Korda M, Hurák Z.
\newblock Controlled synchronization of coupled pendulums by {Koopman} {Model}
  {Predictive} {Control}.
\newblock Control Engineering Practice. 2023 Oct;139:105629.

\bibitem{budisic_applied_2012}
Budišić M, Mohr RM, Mezić I.
\newblock Applied {Koopmanism}.
\newblock Chaos: An Interdisciplinary Journal of Nonlinear Science. 2012
  Dec;22(4):047510.

\bibitem{nuske_finite-data_2022}
Nüske F, Peitz S, Philipp F, Schaller M, Worthmann K.
\newblock Finite-{Data} {Error} {Bounds} for {Koopman}-{Based} {Prediction} and
  {Control}.
\newblock Journal of Nonlinear Science. 2022 Nov;33(1):14.

\bibitem{peitz_data-driven_2020}
Peitz S, Otto SE, Rowley CW.
\newblock Data-{Driven} {Model} {Predictive} {Control} using {Interpolated}
  {Koopman} {Generators}.
\newblock SIAM Journal on Applied Dynamical Systems. 2020 Jan;19(3):2162--2193.
\newblock Publisher: Society for Industrial and Applied Mathematics.

\bibitem{korda_convergence_2018}
Korda M, Mezić I.
\newblock On {Convergence} of {Extended} {Dynamic} {Mode} {Decomposition} to
  the {Koopman} {Operator}.
\newblock Journal of Nonlinear Science. 2018 Apr;28(2):687--710.

\bibitem{mamakoukas_local_2019}
Mamakoukas G, Castano M, Tan X, Murphey T.
\newblock Local {Koopman} {Operators} for {Data}-{Driven} {Control} of
  {Robotic} {Systems}.
\newblock Robotics: science and systems. 2019 Jun.

\bibitem{korda_optimal_2020}
Korda M, Mezić I.
\newblock Optimal {Construction} of {Koopman} {Eigenfunctions} for {Prediction}
  and {Control}.
\newblock IEEE Transactions on Automatic Control. 2020 Dec;65(12):5114--5129.

\bibitem{lusch_deep_2018}
Lusch B, Kutz JN, Brunton SL.
\newblock Deep learning for universal linear embeddings of nonlinear dynamics.
\newblock Nature Communications. 2018 Nov;9(1):4950.
\newblock Number: 1 Publisher: Nature Publishing Group.

\bibitem{forssell_closed-loop_1999}
Forssell U, Ljung L.
\newblock Closed-loop identification revisited.
\newblock Automatica. 1999 Jul;35(7):1215--1241.

\bibitem{brunton_data-driven_2022}
Brunton SL, Kutz JN.
\newblock Data-{Driven} {Science} and {Engineering}: {Machine} {Learning},
  {Dynamical} {Systems}, and {Control}.
\newblock Cambridge University Press; 2022.
\newblock Google-Books-ID: rxNkEAAAQBAJ.

\bibitem{kim_dynamic_2015}
Kim S, Kwon S.
\newblock Dynamic modeling of a two-wheeled inverted pendulum balancing mobile
  robot.
\newblock International Journal of Control, Automation and Systems. 2015
  Aug;13(4):926--933.

\end{thebibliography}

\end{document}